# Evolution of macromolecular structure: a 'double tale' of biological accretion

Derek Caetano-Anollés[1], Kelsey Caetano-Anollés[2] and Gustavo Caetano-Anollés[3]

[1]*Abteilung Evolutionsgenetik, Max-Planck-Institut für Evolutionsbiologie, Plön, Germany;* [3]*Department of Agricultural Biotechnology, and Research Institute for Agriculture and Life Sciences, Seoul National University, Seoul, Republic of Korea;* and [3]*Department of Crop Sciences, University of Illinois, Urbana, Illinois, USA.*

**Abstract** The evolution of structure in biology is driven by accretion and change. Accretion brings together disparate parts to form bigger wholes. Change provides opportunities for growth and innovation. Here we review patterns and processes that are responsible for a 'double tale' of evolutionary accretion at various levels of complexity, from proteins and nucleic acids to high-rise building structures in cities. Parts are at first weakly linked and associate variously. As they diversify, they compete with each other and are selected for performance. The emerging interactions constrain their structure and associations. This causes parts to self-organize into modules with tight linkage. In a second phase, variants of the modules evolve and become new parts for a new generative cycle of higher-level organization. Evolutionary genomics and network biology support the 'double tale' of structural module creation and validate an evolutionary principle of maximum abundance that drives the gain and loss of modules.

Derek Caetano-Anollés is Research Fellow at the Department of Evolutionary Genetics at the Max-Planck Institute for Evolutionary Biology, Plön, Germany. He holds a PhD in Cell and Developmental Biology from the University of Illinois, Urbana-Champaign. His current research interests include the transcriptional effects that govern neurogenesis as well as the genetic and molecular factors that affect facial morphology during evolution.

Kelsey Caetano-Anollés is Research Scientist at the Department of Agricultural Biotechnology and Research Institute for Agriculture and Life Sciences of Seoul National University, Republic of Korea. She holds a PhD in Animal Sciences from the University of Illinois, Urbana-Champaign. Her current research interests include application of bioinformatics tools and population genetics principles on analysis of genomic polymorphisms, specifically on positive traits in livestock.

Gustavo Caetano-Anollés is Professor of Bioinformatics and University Scholar of the Department of Crop Sciences and Affiliate of the C. R. Woese Institute for Genomic Biology at the University of Illinois, Urbana-Champaign. He holds a PhD in Biochemistry from the University of La Plata in Argentina and is interested in creative ways to mine, visualize and integrate data from structural and functional genomic research, focusing on evolution of molecular structure and networks in biology. His group is also interested in the spontaneous mutation process and in understanding genetic diversity. Correspondence should be addressed to Dr. Gustavo Caetano-Anollés, Department of Crop Sciences, University of Illinois, 332 NSRC, 1101 West Peabody Drive, Urbana, Illinois, 61801, USA. E-mail: gca@illinois.edu

**Introduction**

> *"A double tale I'll tell. At one time one thing grew to be just one*
> *from many, at another many grew from one to be apart.*
> *Double the birth of mortal things, and double their demise."*
>
> – Empedocles, *On Nature, P. Strasb. Gr. Inv.* 1665-6, lines 233-235[1]

Parts often form bigger wholes, which can combine with other wholes in an ever-expanding process of accretion. This process provides opportunities for growth and innovation. Accretion manifests at different timeframes and appears universal. Physical cosmology offers a number of examples describing how the gradual accumulation of matter into massive objects occurs at varying levels of granularity. Planets form by accretion of dust from giant molecular clouds left behind during the formation of stars[2]. Interactions of dust grains in the nebula accrete into planetary disks that orbit around the central proto-star to form planets. Similarly, the gravitational collapse of giant molecular clouds and dense interstellar media form circumstellar disks of orbiting matter that spiral inward towards growing central stellar bodies to form stars[3,4]. At even larger levels, the gravitational collapse of vast interstellar clouds of dust and gas form a variety of stellar masses by processes of clumping and merging[5], including evolving quasars, galaxies, clusters and superclusters[6].

Unsurprisingly, life and its consequences are the subject of accretion. Molecular components become parts of growing molecules and macromolecules, which also interact and merge with other growing molecules and macromolecules to form molecular complexes that make up higher levels of molecular and cellular structure[7]. For example, we used phylogenetic methods to trace the evolutionary growth of the ribosome, the molecular complex responsible for protein synthesis in the cell[8-10]. Figure 1A shows a timeline describing how ribosomal RNA (rRNA) helices and ribosomal proteins (r-proteins) accrete in evolution to form the modern ribosomal biosynthetic complex. Similarly, cells accrete into more complex cellular and organismal assemblies, including cellular consortia, multicellular organisms, and populations of diversifying individuals. Organismal assemblies can be massive. Eusocial communities, as we see with ant colonies, can unite into super-colonies of billions of individuals. For example, a super-colony of Argentine ants was detected that spanned the Mediterranean and Atlantic coasts of Southern Europe[11]. This super-colony became a global mega-colony that took over the world when it recently fused with Japanese and Californian counterparts[12].

The products of animal intelligence also accrete. Cities, the settlements of humankind, are often the subject of inexorable processes of growth triggered by population growth, innovation and wealth creation[13]. Urban organization and dynamics appear linked to knowledge creation and economies of scale that guarantee sustainable development. Figure 1B summarizes the outcome of a 180-year timeline of population growth and urbanization of the Manhattan Island, the most densely populated borough of New York[14]. The process of growth started in Lower Manhattan (which currently holds the financial district), but then spread northward and more evenly. The accretion of high-rise buildings in the financial district emerged in the early 1900's as a consequence of increasing population and wealth in the region (Figure 1B). Thus, the development of higher and higher levels of organization result in more complex structure that is permanently growing and diversifying. Here we focus on the processes of growth, modularity, and innovation that exist in the biological world.

**Structures in 4D and the rise of modules**

The 3-dimensional (3D) atomic structure of macromolecules can be effectively described as a nested hierarchy of *modules*[7]. Modules are sets of integrated parts that cooperate to perform a task and interact more extensively with each other than with other parts and modules of the system[15]. While the cooperative performance of specific functions distinguishes modules from each other, the rise of modularity in structural biology can be seen as a 'nucleation' process, i.e. the emergence of cohesion in selected parts of the makeup of a molecular system. Cohesion refers here to the dynamical stabilities of component parts of the system when their dynamics are constrained by the system as a whole. These stabilities ultimately determine if the system can be easily decomposed into parts[16].

The 3D structure of functional nucleic acids and proteins contains a variety of structural modules, which we use in the study of nucleic acid and protein evolution[10]. The structure of functional RNA is mostly determined by A-form helical elements of structure known as *'stems'*. These modular elements arise when H-bonding interactions between nucleotides are optimally satisfied during base pairing. Note that H-bonding occurs between the identity-conferring elements of the repeating units of the nucleic acid polymer. Initial helix-forming interactions are further and crucially stabilized by base pair stacking interactions. Stems connect with each other through *'junctions'*, branching points of the molecules that are largely unpaired and form multi-branched loops. A typical tRNA molecule contains four helical 'arms' and a single 4-way junction, which twist the molecule into an L-shape 3D structure. Similarly, the small 5S rRNA molecule contains three arms with 5 stems and a central 3-way junction that twist the branched structure. In contrast, the small and large subunits of the structural core of the ribosome, which is universal, typically contain 50 and 100 stems and 18 and 29 junctions, respectively. These large arrangements of ribosomal stems and junctions give rise to massive globular conglomerates that distill 3.3 billion years (Gy) of history. Figure 2A illustrates RNA modules with the atomic structure of a typical stem and the central junction of 5S rRNA from *Escherichia coli*. The structure of a protein is much more flexible and fine grained. Proteins fold and organize into structured and unstructured regions. The structured regions are mostly stabilized by helical and strand elements of secondary structure, which form when H-bonding interactions are established within the protein polypeptide backbone. Note that in contrast with nucleic acids, these H-bonding interactions do not involve the identity-conferring elements of the repeating units of the polypeptide. This difference has notable consequences for the energetics of folding. Elements of secondary structure come together to form super-secondary motifs that are ~25-30 amino acid residues long, initially in the form of *'loops'*. Loops are returns of the polypeptide chain that are often stabilized by interatomic attractive forces (van der Waal forces, dense packing) that lock their conformations. These loops make recurrent super-secondary structures, which define the topology of each of 1,221 folds (classified in SCOPe 2.06) that currently describe the structure of protein *'domains'*. Domains are higher level modules of compact structure, molecular function and structural evolution. They combine with other domains to form multi-domain proteins. Figure 2B illustrates the hierarchical embedding of modules in the structure of the D-ribulose-1,5-biphosphate carboxylase/oxygenase (RubisCO) enzyme[17]. RubisCO is part the most important carbon fixation pathway of metabolism and is the most abundant enzyme on Earth.

Over the last two decades we have developed phylogenomic methods that turn the 3D view of macromolecular structure into a 4-dimensional (4D) structural view that incorporates the axis of time (reviewed in[7,10,18]). Macromolecular accretion and change depend on genomic changes (point mutations, insertions, deletions, rearrangements, fusions, and fissions) that promote molecular innovations and benefit organismal persistence. These changes leave behind features embedded in structure that have historical memory. We use standard cladistic approaches to build phylogenetic trees from a census of structures that are encoded in thousands of genomes. The 3D-to-4D strategy reconstructs the past by analyzing structural data with evolutionary models capable of producing tree-like statements of history, which are known as phylogenies. Standard phylogenies have the form of phylogenetic trees, but can also be represented with phylogenetic networks. Our phylogenomic methods revealed two opposing processes controlling the evolution of nucleic acids and proteins (Figure 3A). One grows individual modules from lower-level parts, *'to be just one from many'*, unifying parts into wholes. The other grows the emerging wholes into many, *'from one to be apart'*, diversifying these modules into novel parts. Nucleotides are monomers of polynucleotide chains that collapse to form helical structures. These RNA structures assemble into more complex 3D arrangements through condensations, which then assemble into junctions to build up larger molecular complexes. Similarly, amino acids condense into dipeptides that act as modules for polypeptides. These short molecules fold to form loops, which then act as modules to build the structural domains of proteins.

The *'double tale'* of module growth and diversification is one of competitive optimization, one that *'doubles'* the *'birth'* and *'demise'* of the emerging modules. We have put forth a theoretical framework that explains module creation through a biphasic process of diversification[19]. We illustrate the framework with the evolution of a network in which parts are nodes and connections are linkages of parts in the molecular system (Figure 3B). In a first phase, parts are at first weakly linked and associate variously. As they diversify, parts compete with each other and are selected for performance. The emerging interactions constrain their structure and associations. This causes parts to self-organize into modules with tight linkage, which are dynamically resistant to fluctuations and change. In the second phase, variants of the modules evolve and become new parts for a new generative cycle of higher-level organization. The timeline of networks presented in Figure 3B describes the dynamics of interacting linkages between parts of a growing system. This timeline shows the rise of hierarchical modularity in the structure of the evolving network.

**Accretion of RNA structure**

We have shown that RNA molecules carry deep phylogenetic history in the topology and thermodynamics of their 3D structures (previously reviewed in this journal[18]). The dynamics of RNA folding is linked to the secondary structure of RNA and is negatively correlated with chain length[20,21]. Structures collapse into few stable conformations when they quickly reach local folding solutions that are evolutionarily optimized[22]. These conformations foster helical structural modules compatible with the length and history of the molecules. This history can be mined with phylogenomic methods that build trees of molecules (ToMs) and trees of substructures (ToS) from features describing the topology and thermodynamics of helical stem and non-paired segments in RNA conformations. ToMs and ToSs are data-driven models of the history of the molecular system and its component parts, respectively, which they label at their leaves. Trees are rooted and therefore capable of defining an 'arrow of time'. They also make explicit the first and second 'tale' of accretion. The balanced topologies of ToMs make them useful to study the diversification of molecules and the cellular world. The comb-like topologies of ToDs allow building timelines of appearance of parts in molecules. We used these 'natural histories' to study the origin and evolution of the most ancient RNA molecules, including tRNA[23,24], RNase P RNA[25] and the large and small rRNA subunits[8-10] and 5S rRNA[26].

*Accretion of RNA helical stems to form complex RNA structure.* ToMs and ToDs unravel macromolecular accretion. To illustrate their power, we quickly review an analysis of the evolutionary history of 5S rRNA[26]. Figure 4A shows a ToM that uses the 5S rRNA molecules as organismal proxies to describe a tree of cellular life, which we find is rooted in Archaea. The placement of Archaea at the base of the tree suggests its early diversification from a stem line of descent. This scenario is compatible with considerable other evidence, including phylogenomic data recovered from the structure of RNA, structural domains in proteomes (the entire protein repertoire of an organism), and Gene Ontology definitions of molecular functions[27]. ToDs revealed consistent patterns of origin and diversification of 5S rRNA structure when the entire molecular set or molecular sets corresponding to either Archaea, Bacteria or Eukarya were examined (Figure 4B). First helix S1 was found to be the most ancient stem of the molecule regardless of the dataset examined. Originally, it was probably a hairpin structure similar to the one described in the structural model of Figure 2A. Second, the oldest S1 stem was consistently found historically linked to the 5' terminal free end, while the 3' end was younger. This same ancestral link to the 5' unpaired terminal region was also identified in tRNA[23] and matches statistical studies of RNA structures[28]. This suggests a possible early functional association of free ends of the ancestral molecule with tRNA CCA-adding enzymes or a common origin of tRNA and 5S rRNA. Third, weak G:U base pair interactions appeared late in evolution first in association with the S4 stem and then with S1. These motifs help stack guanosines when they are in tandem and stabilize interactions with water. Fourth, we found the central ribosomal junctions that branched the molecule formed for the first time in Archaea when the primordial S1 and S3 stems added the branching S5 stem. The delayed formation of the branched structure in Bacteria and Eukarya lends further support to the archaeal rooting of the tree of life. Thus, the 5S rRNA molecule diversified as the living world was establishing its first modern microbial lineages. We also examined the history of r-proteins associated with the molecule that was inferred from a census of structural domains in hundreds of genomes. Remarkably, we found that the RNA molecule coevolved with the r-proteins, staring with the very early association of the S1 stem with the translation protein SH3-like domain present in the L21e r-protein, which appeared ~2.9 Gya according to a clock of folds[29]. This time coincides with the rise of aerobic metabolism and atmospheric oxygen in our planet. Coevolution is here defined as the coordinated succession of structural changes mutually induced by the increasingly interacting and growing RNA and protein molecules. Since the ribosome made its debut 3.3 Gya (Figure 1A), the ribosomal RNA-protein co-evolutionary patterns confirm the late addition of the 5S rRNA molecule to the ribosomal ensemble. Two crucial observation further confirm its late arrival. 5S rRNA is the only rRNA molecule that binds r-proteins before assembly into the ribosome[30]. Furthermore, 5S rRNA binding to the large ribosomal subunit depends on extensive interactions with proteins[31], except for few RNA–RNA interactions with the S4 and S5 stems that we found are evolutionarily derived.

*Molecular accretion and the formation of ribonucleoprotein complexes.* Harish and Caetano-Anollés[9] extended preliminary work[8] by embarking on a historical analysis of the entire ribosomal complex. The study confirmed the coevolutionary history of ribosomal RNA and proteins. While it validated initial findings drawn from the 5S rRNA molecule[26], it also explained the full-fledged complexity of ribosomal structure (reviewed in Caetano-Anollés and Caetano-Anollés[10]). Phylogenomic analysis of rRNA and r-protein structures of the small and large ribosomal subunits uncovered an evolutionary timeline of accretion. This timeline described the evolution of the universally conserved ribosomal complex, which was visualized by coloring relative evolutionary age on 3D atomic ribosomal models (Figure 1A). The study found a tight correlation between the age of rRNA stems and interacting domains of r-proteins as they co-evolved to form a fully functional ribosomal core. The oldest protein (S12, S17, S9, L3) appeared concurrently with the oldest rRNA substructures responsible for decoding and ribosomal dynamics. This occurred 3.3–3.4 Gya according to the clock of folds. These RNA structures included the central ribosomal ratchet and two hinges of the small subunit and the L1 and L7/L12 stalks of the large subunit. All of these structures are important for ribosomal movement of tRNA in the complex. During the very early steps of ribosomal evolution, RNA substructures appeared at first in orderly fashion until the formation of a 5-way and 10-way junctions in the small and large subunits, respectively, at which point a '*major transition*' in ribosomal evolution occurred 2.8–3.1 Gya. This transition brought ribosomal subunits together through inter-subunit bridge contacts and stabilized loosely evolving ribosomal parts. It also developed structures supporting interactions with tRNA and a fully-fledged peptidyl transferase center (PTC) with exit pore capable of protein biosynthesis. A '*second transition*' occurred later concurrently with the Great Oxygenation Event of our planet (~2.4 Gya). This occurred when the L7/12 protein complex that stimulates the GTPase activity of EF-G and enhances ribosomal efficiency was accreted into the ribosome. Coupling the evolutionary timelines of tRNA and rRNA structure with annotations of their interactions revealed that the tRNA cloverleaf structure was already a fully formed when the PTC appeared in evolution[24]. Thus, fully formed tRNA molecules played other roles before being recruited for processive protein biosynthesis, perhaps as cofactors of peptide-producing dipeptidases and ligases. See Caetano-Anollés et al.[32] for a more detailed elaboration of a data-driven hypothesis for the origin of translation and genetics. In summary, protein-RNA coevolution manifests throughout the timeline of ribosomal accretion. This observation strongly supports the origination of the ribosome in a molecular world that contained both evolving proteins and RNA, which challenges the ancient 'RNA world' hypothesis.

Recently, a number of tRNA remote homologies (similarities believed to arise from common ancestry) were detected in the small and large subunits of rRNA, which suggested that the ribosome was built piecemeal from primordial tRNA-like molecules[33]. These and earlier results that suggest the PTC originated in tRNA[34] added evidence to the proposal of David Bloch and colleagues in the 80s that both tRNA and the ribosome had a common remote evolutionary origin[35]. Such a proposal has significant implications. It suggests that the ribosome grew by recruitment from a multiplicity of molecular parts that were perhaps already acting as amino acid acceptors[36]. The significant historical patchwork observed when the age of helices and proteins was colored onto the ribosomal 3D model (Figure 1A) provides phylogenomic support to this recruitment hypothesis. In fact, a study of putative branch-to-trunk insertions in ribosomal junctions reveals at least 5 and 14 possible independent origins for the small and large subunits[37]. Thus, the ribosome could have originally consisted of ~20 different rRNA molecules loosely linked together. tRNA sequences in rRNA also showed remote homologies to elongation factors, synthetases, RNA polymerases and nucleotide biosynthetic enzymes[33], all of which suggest the tRNA-like molecules also assembled to form primordial genomes.

**Accretion of protein structure**

Profile hidden Markov models (HMMs) of structural recognition assign folds of known structural domains to protein sequences with high accuracy and low error rate. For example, we used statistical models developed in HMMER3[38] to generate a structural census of structural domains. Domains were defined at various levels of protein structural abstraction in both SCOP[39] and CATH[40], the gold standards of domain classification. The proteomic occurrence and abundance of all known domain structures was computed for each proteome across a wide range of cellular organisms and viruses. This information, which embodies millions of domains, was used to build rooted phylogenomic trees of domains (ToDs) and trees of proteomes (ToP). Since domains are structural, functional and evolutionary parts of proteins in proteomes, the comb-like ToDs describe the evolution of the protein world and can produce chronologies of first appearances of domains. A clock of fold links these chronologies to the geological record[29]. Similarly, since proteomes represent protein repertoires that are so central to life, they can be considered proxies for cellular organisms or viruses. Thus, ToPs become bonafide 'trees of life'.

The first study involved an analysis of 32 microbial organisms that had been sequenced at that time[41]. Recent analyses extended the approach to thousands of cellular organisms and viruses[42]. Figure 5 shows an example ToD describing the evolution of second highest hierarchy of CATH. It provides a hypothesis of origin and diversification of protein structural topologies, i.e. a hypothesis of history of the protein origami[41,43]. The most ancient architectures were sandwiches and bundles, and the most derived were highly symmetric structures such as propellers and prisms. Chronologies of domain innovation directly inferred from ToDs are very useful. They have been used for example to trace the origin and evolution of metabolic networks[44]. Figure 6 shows networks describing the early evolution of purine metabolism, the oldest subnetwork of metabolism[45]. Metabolic subnetworks of these kinds are considered modules[46]. The figure shows how the metabolic module holding the purine biosynthetic pathway assembled as a patchwork through processes of recruitment in the unification-diversification process that we described in Figure 3. Nodes of the subnetwork represent metabolites. Transformation to other metabolites via de activity of enzymes define interaction links, which are generally represented by directed connections (arrows) in the network. The graph is at first fragmented into small submodules. However, these fragments finally unify into the fully functional, cohesive and modern core of purine metabolic pathways that appeared ~3 Gya. Note that the fragmented submodules are always unified by a network established at lower hierarchical level of interaction. In our case this unifying network is of prebiotic non-enzymatic reactions that are inferred from current chemical experimentation. These unifying links are given as red lines (sometimes dashed) in the evolving networks. As prebiotic reactions of early Earth are replaced by modern enzymatic counterparts, the different submodules of modern metabolism coalesce into the main nucleotide interconversion (INT), metabolism/salvage (CAT), and biosynthetic (BIO) pathways that exist today[45]. Further analytic explorations provided additional evidence supporting the validity and predictions of the biphasic model of Figure 3C:

(i) *Emergence of hierarchical modularity in metabolic networks:* Network modules can be detected with a variety of approaches. One example is through the calculation of a clustering coefficient—a metric that describes the average probability that two neighbors of a node are also connected—which can be used as a measure of modularity[47]. The coefficient scales negatively with the number of connections when networks are hierarchically modular. Alternatively, sets of nodes that are densely connected with each other are said to form a community. Communities can be detected with for example hierarchical clustering methods, the path-pruning Girvan-Newman algorithm, or the maximization of modularity functions. For example, the statistical analysis of modularity with the Louvain maximization method reveals the rise of community structure in the evolving purine metabolic subnetworks (Figure 6). We have extended these kinds of analyses to the entire metabolic network (F. Mughal and G. Caetano-Anollés, ms. in preparation). Increases in clustering coefficients, modularity and community structure confirm the rise of modules in evolution of all known metabolic networks.

(ii) *Emergence of protein structure by accretion of amino acids and dipeptides:* We studied the dipeptide make-up of protein domains and their significance for both the genetic code and the structure of proteins[32]. We generated a census of the 20+ possible amino acids and the 400+ possible dipeptides and studied if their distribution along the timeline of domains was statistically biased. The dipeptide composition of domains along the evolutionary timeline showed a biphasic pattern of dipeptide use, with early and late domains showing higher levels of dipeptide representation in the fold structures. This pattern is compatible with our biphasic model. Remarkably, we found that dipeptides of ancient domains appearing prior to genetics were enriched in hydrophobic amino acids and that their participation in flexible loop regions was underrepresented. Findings suggest an early association of protein with membranes and confirmed previous phylogenomic studies of folding speed that show protein flexibility is an important evolutionary driver for the structure of proteins[48].

(iii) *Emergence of domains by accretion of loops:* The functional diversity of proteins likely emerged from the rearrangement of super-secondary structural building blocks made of helix, strand and turn segments (e.g. αα-hairpins, ββ-hairpins, βαβ-elements). Some recent studies identified non-combinable[49] and combinable[50] loop motifs responsible for very early molecular functions. We traced the combination of loops in domains by generating networks of loops and domains and tracing the evolution of these networks along the domain timelines[51]. We found that both modularity and community structure were emergent properties of the evolving network when we used metrics of connection density, hierarchy, and modularity of network structure. We also uncovered evolutionary patterns of innovation that matched the geological record[17].

(iv) *Emergence of multi-domain proteins by combination of protein domains:* Protein domains combine in evolution to produce a substantial number of multi-domain proteins, generally involving 26-32% of domains and 58-83% of proteins of a proteome[52,53]. These domains are either repeated or combined with other domains in defined order as one travels along the polypeptide chain. We generated ToDs that describe the evolution of both domains and domain combinations and produce timelines of protein appearance in evolution[52]. Remarkably, timelines revealed the early rise of single-domain proteins in evolution, with structures that were generally multifunctional. This gradual rise was followed by an explosive increase of domain combinations, a 'big bang', that occurred during the rise of eukaryotic organisms. The modular combination of domains in proteins that dominates the protein world is the product of a number of biological activities, including the chromosomal recombination, retrotransposons, intronic rearrangement of domain-encoding exons and faulty excision of introns. These processes result in fusions and fissions that combine, recruit, and split domains in proteins. Fusions and fissions were traced along the timeline. Fusions occurred early and their frequency rose gradually to domination. In contrast, fission processes produced multi-functional proteins quite late in evolution. We have generated evolving networks that describe the combination of domains in proteins (M.F. Aziz and G. Caetano-Anollés, ms. in preparation), revealing again the rise of modularity in network structure. This suggests the existence of higher levels of modules emerging in the combinatorial interplay of protein domains.

**Gain and loss**

There is significant evidence of gradual growth in the accretion process. In proteins, accretion complies with the '*principle of spatiotemporal continuity*' of Leibniz and materializes in allometric scaling relationships compatible with the Heaps law, a law that describes the growth and evolution of vocabularies through correlations of language properties with either time or accumulating innovation[54]. The existence of growth in the macromolecular world does not imply that loss does not occur. Both gains and losses are universal occurrences manifesting at widely different levels of biological complexity. Losses however are more difficult to track in evolution. To illustrate, the 'twin towers' of the World Trade Center are not present in the historical record of Figure 1B since they were destroyed in the September 11, 2001 terrorist attack on the City of New York, and the simulation shows only the history of buildings that exist today. The twin towers were the tallest building in the world (417 m) during 1972-1973 until they were overshadowed by the Sears Tower of Chicago that was built in 1974. Their 'tubular' cantilevered design was an example of engineering innovation in the construction of the 'supertall' buildings (>300 m in height) of their time.

Similarly, it is highly likely that many domain innovations were lost or replaced, and that many never fully materialized in the world of proteins. Many of these instances are likely impossible to detect. In this regard, we have studied global patterns of domain gain and loss in the three superkingdoms of life[55]. We reconstructed ToPs and retraced the history of changes of domain occurrence and abundance in proteomes along their branches. While both domain gains and losses were frequent events in the evolution of cells, domain gains always overshadowed the number of losses (Figure 7). We also found that domain gain-to-loss ratios increased in evolution, supporting an evolutionary acceleration of growth. Observations from the phylogenomic reconstructions are highly significant and support the role of competitive optimization in the generation of modules. They highlight Empedocles' adagio of birth and demise that already reflected Darwinian thinking 24 centuries ago. However, they also show a pervasive accumulation of innovations in biology, which we will now discuss.

**A succession of increasingly complex modular parts**

The biphasic model, which we illustrated with the network paradigm of Figure 3B, embodies an important property of modules: the fractal-like embedding of modules onto modules. This property subsumes the hierarchical modularity of networks described in Figure 6. A fractal is an 'evolving symmetry' that arises when multi-level dependencies appear in a hierarchy. Phylogenomic analyses uncovers the existence of a succession of increasingly complex modules nested in the hierarchy of evolving nucleic acids and proteins (Figure 8). These modules always embed lower level modular parts. For example, ribosomal growth implies an evolutionary accumulation of parts making up a nucleic acid scaffold onto which a multitude of other molecules (e.g. r-proteins) and interactions (e.g. A-minor interactions) gradually accrete[10]. An almost linear dependency of the number of nucleotides in paired and unpaired regions of the rRNA molecules against the age of helical segments reveals a flattened S-shape scaling relationship that models the RNA growth of the ribosomal complex (Figure 8A). Remarkably, the first appearance of a 3-way junction, a larger and more complex module than a nucleotide monomer, occurred for the first time exactly during the first major transition of ribosomal evolution that brought together the small and large subunits for protein biosynthesis[9]. Their accumulation unfolded later with an S-shaped scaling relationship. The successive accumulation of modules is even more evident in proteins. Two successive 'waves' of gain and loss of protein domains are clearly evident in the phylogenomic reconstruction exercise described in Figure 7. One appeared before and the other after the rise of the fully diversified organismal world. When the history of domains and domains combination is traced in the evolutionary timeline an early rise of single domain proteins is followed by a massive 'big bag' of domain combinations[52], showing that the combination of protein modules was a late evolutionary development (Figure 8B). Remarkably, this same emergent behavior occurs in cities.

Individuals, colonies, or communities of organisms often build enclosed structures that help them survive and reproduce. These evolutionary adaptations provide important benefits. Microbes sometimes grow on surfaces or sediments protected by pericellular matrices. Plants and fungi form buried or aerial structures for vegetative reproduction and dispersal. Vertebrates and some invertebrates construct shells, nests and hives. Even viruses produce capsids, proteinaceous shells for effective dispersal. Humans gather in settlements protected by building structures for shelter, storage and work. Settlement communities embody social units organized in hamlets, villages, towns and cities. These settlements can be readily identified by housing, utility and transportation infrastructure. Housing infrastructure include buildings, which often serve as dwellings. They tend to be proportional to human population. Since settlements and buildings are the products of human endeavor, their makeup is likely genetically encoded and hierarchically modular. Indeed, the settlement history of the Manhattan Island, which began with the Dutch in 1609, shows a clear succession of increasingly complex modular parts, which ultimately gave rise to the modern City of New York (Figure 8C). Human population divides into dwellings in analogy to how amino acids apportion into protein structures. Social units (including families) live in single or multistory housing modules resembling single domain or multi-domain proteins, respectively. The simple exercise of tracing the history of human population, population densities and building 'floor area ratio' (FAR) in the island[14] shows the emergence of high-rise building structures and a strong tendency toward vertical depth (Figure 8C). This tendency accommodates the activities or expanding human population. FAR measures the ratio of a building's floor area and the area of the plot in which it is built. High-rise buildings have high FAR and on average increase this index of vertical depth. Manhattan's population and population densities increased steadily until the start of 1900s[14]. Growth of the built-up area could not keep up with rapid population growth, which was triggered mainly by the steep rise of trade in New York harbor. In the 1850s, this trade represented ~70% of the total trade of the country. Economy triggered population growth and that triggered the need for building expansion through increased FAR. This need pushed both technological advances in high-rise building construction (e.g. steel frames, elevators) and urban expansion through transportation infrastructure (e.g. trains, highways). The increase of vertical depth in the city, now a worldwide phenomenon of large cities, coupled to urban expansion, impacted the distribution of the rising population. Both population growth and density decreased in the island after the annexation of Brooklyn, Queens, Bronx and Richmond County in 1898 and the building of a subway transport system in 1904[14]. These developments enabled urban expansion and massive commuting, which led to an expanding transient (on-the-go) population. Today, Manhattan's population doubles every day from 1.6 to 3.1 million due to commuters that travel from other boroughs to work in the inner city. The history of New York therefore embodies a remarkable succession of module development. First, the unique deep ice-free harbor of the Dutch settlement seeded a port, factories and trade, which attracted population and commerce through competitive optimization. A hub-and-spoke transportation-trading system later on increased the wealth of the entire region and started to consolidate a financial module in the city that had already significant national impact in the 1850s and later made New York the world's financial capital. Second, the initial financial preeminence of the city jumpstarted a massive rise of population (the lower level city modules), which peaked in 1910. This massive rise resulted in crowding, which fueled the rise of a new kind of use of the housing module throughout the island, the high-rise building. Third, these buildings diversified into supertall buildings, especially in the now functionally specialized financial district module of Lower Manhattan (Figure 1B). Finally, this trend towards district and building modularity became a planetary tendency of vertical depth when the modular patterns of New York were recruited into other cities. This is illustrated by the historical timeline of the tallest buildings and the number of the supertall building of the worlds (Figure 8D). Ten of the 15 tallest buildings in history were built in New York and only appeared elsewhere in the world in 1996.

The worldwide recruitment of high-rise modules has been slow. Buildings over 100 m tall did not appear in Europe until after the 1950s and rarely exceeded 250 m in height[56]. A systematic analysis of the shape and structure of European skyscrapers showed considerable diversification around a central core with similar floor plan[56], revealing a second phase of high-rise building emergence of the 'double tale' model. This diversification could be apportioned into 4 categories of symmetry: uniaxial, biaxial, multiaxial, and irregular, which showed remarkable similarities with the classification of the symmetry of protein complexes (Figure 9). Symmetry is the concept of the repetitive arrangements of similar modular objects in space by the use of basic operations of rearrangement, including translation, rotation, inversion and gliding. Protein multimers can be arranged in cyclic (C), dihedral (D), higher cubic (O, T and I) and helical symmetries (H)[57]. These symmetries match floor plans of the high-rise buildings (Figure 9). One significant difference however is the strong vertical directive of the building versus a number of possible axes of growth in protein complexes.

**Diversification as prelude to innovation**

The gain and loss of structural domains in proteome evolution illustrated by the growth curves of Figure 7 reveal a clear succession of sequential logistic S-shaped wavelets ('loglets'). These sigmoid curves of growth are logistic components of diffusion and substitution that model time-series data[57]. They have been used for example to model the rise and fall of infrastructures, including transportation infrastructures[58]. However, they also apply to biology. Lotka and Volterra in the 1920s described a process of competitive optimization that led to substitution patterns in a population subjected to constraints, limiting factors such as scarcity of matter, energy, space or information. While the original substitution model involved the replacement of species in a population through 'lock-in' patterns, the overall concept can be used to model the competitive optimization of innovation variants that lead to the generation of modules. Simple symmetrical logistic models can be complemented with non-symmetrical or skewed growth models or generalized logistic functions describing the gradual takeover of diversified variants[57,58]. Decompositions through the Fisher-Pry transform normalize the logistic curves to straight lines helping untangle the complex growth patterns that are expected when analyzing the hierarchical complexity of modules.

S-curves can be explained with causal models of growth and diffusion of innovations. Diffusion models were originally developed to describe the spread, adoption and effects of innovations in society, in particular the spread of cultural innovations from one society to another in socio-economic systems[59]. However, diffusion is a general process that 'communicates' innovations over time in a system. Communication here refers to the transfer and 'adoption' of a design, function or concept to other parts of a system (e.g. organisms in a population). Double S-curves are typical of paths of high performance in diffusion of innovation models. These bi-logistic curves (wavelets) often overlap in time with different magnitude and with different speeds generating sequential patterns of loglet superposition, convergence and divergence[57]. They are illustrated in Figure 10. When innovations are communicated among entities over time, some of them are adopted in the system initiating a 5-step decision-making innovation lifecycle that involves 'innovators', 'early adopters' (trendsetters), an 'early majority', a 'late majority', and 'laggards', in that order. One of the most general and popular 'communication' model of these types is the Bass method[60]. The method considers a homogeneous population with a fraction of innovation adopters ($\rho$) increasing when they meet non-adopters (1- $\rho$) under intrinsic forces with rate *q* (coefficient of *imitation*) and under external forces with rate *p* (coefficient of *innovation*). Equation 1 and its derivative, Equation 2 (Figure 10), describe the adoption process. Note that when *p* = 0, the Bass model becomes isomorphic with the logistic model and only imitation by non-adopters governs the diffusion process. Similarly, when *q* = 0, the Bass model reduces to the exponential distribution in which emergence of innovation drives change.

In molecular biology, innovation communication involves mutational diffusion in sequence and structure space when organisms select their best molecules in response to internal and external needs. When pairs of loglets induce biphasic patterns of growth, these bi-logistic wavelets describe complexity in the emergence of modules and innovation. This complexity can be decomposed into logistic components. For example, the flattened S-shaped scaling of nucleotide number in helical stems of rRNA also embeds wavelets (Figure 8A). One wavelet appears before and another after the first major transition of ribosomal evolution[9]. This is caused by subsets of nucleotides making up higher level modules, such as junctions, A-minor and r-protein interactions, and even higher level structural interactions. Similarly, the combinatorial rise of multi-domain proteins involves a wavelet, with its first loglet being of small magnitude appearing before the 'big bang' of the protein world and involved in the discovery of domains structures, and its second loglet being of great magnitude after the 'big bang' and involved in domain-domain interactions (Figure 8B). With cities, the rise of a loglet of human population leads to a loglet of vertical city depth (Figure 8C and D). Similar successive innovation waves can be seen in U.S. music recording media and transportation infrastructure[57].

**Conclusions**

Accretion involves growth and diversification, which cause the evolutionary accumulation of innovations[54]. What drives their generation? We have made the case that information dissipation and modularity pervade biological structure in a way that maximizes energy and information flux through a system[61]. We used Layzer's far-from-equilibrium cosmological model to argue that a simple conservation law links information and entropy. The law states that the sum of possible and instantiated entropy is constant, illustrating two opposing tendencies, one that is dissipative and diversifies and the other that unifies and generates order, structure and complexity. These two opposing tendencies define the 'double tale' of biological accretion when only a minute fraction of a phase space of possibilities (e.g. sequence or structure space) is visited by the system as it wanders in diffusive walks through processes of diversification. These processes of change include mutation, recruitment, and rearrangement. The evolutionary side-product of diffusion and diversification is the emergence of new modules by competitive optimization. The accumulation of successful innovations causes the system to grow and diversify when parts associate or interact to form wholes. This fuels a never-ending accretion process.

Therefore, we propose a '*principle of maximum abundance*' that explains the rise of linkages and modules. This principle pushes systems to grow while maximizing the spread of innovations. The push, however, is limited by constraints imposed by resources and nested cohesive forces that develop in the evolving system. Aggregation is dynamic and involves matter, energy, space, time and information. Linkages involve interactions of many types, from contacts to signals. Emergent modules include modules of structure, function and information.

> *"In the way that many arise as the one again dissolves,*
> *in that respect they come to be and have no life eternal;*
> *but in the way that never do they cease from change continual,*
> *in this respect they live forever in a stable cycle."*
>
> – Empedocles, *On Nature, P. Strasb. Gr. Inv.* 1665-6, *Physica* Book I, lines 241-244[1]

## Acknowledgments

We thank Patrick Lamson-Hall (NYU Stern Urbanization Project) and Jesse Brown (CubeLease) for sharing their data. Research was supported by a National Science Foundation postdoctoral fellowship (award 1523549) to DCA and a grant from the USDA National Institute of Food and Agriculture (Hatch-1014249) to GCA.

## References

1. Janko R (2004) Empedocles, On nature I 233–364: A new reconstruction of p. Strasb. Gr. Inv. 1665–6. *Zeitschrift für papyrologie und epigraphik* Bd., **150**, 1-26
2. Mordasini, C., Molliere, P., Dittkrist, K.-M., Jin, S. and Alibert, Y. (2015) Global models of planet formation and evolution. *Intl. J. Astrobiol.*, **14**, 201–232.
3. Beckwith, S.V.W. and Sargent, A.I. (1996) Circumstellar disks and the search for neighbouring planetary systems. *Nature*, **383**, 139–144.
4. Jofré, P., Das, P., Bertrandpetit, J. and Foley, R. (2017 Cosmic phylogeny: reconstructing the chemical history of the solar neighborhood with an evolutionary tree. *Monthly Notices Roy. Astron. Soc.*, **467**, 1140-1153.
5. Pudritz, R.E. (2002) Clustered star formation and the origin of stellar masses. *Science*, **295**(5552), 68-76.
6. Fraix-Burnet, D., Chattopadhyay, T., Chattopadhyay, A.K., Davoust, E. and Thuillard, M. (2012) A six-parameter space to describe galaxy diversification. *Astronomy Astrophysics*, **545**, A80.
7. Caetano-Anollés, G., Wang, M., Caetano-Anollés, D. and Mittenthal, J.E. (2009) The origin, evolution and structure of the protein world. *Biochem J.*, **417**, 621-637.
8. Caetano-Anollés, G. (2002) Tracing the evolution of RNA structure in ribosomes. *Nucleic Acids Res.*, **30**, 2575–2587.
9. Harish, A. and Caetano-Anollés, G. (2012) Ribosomal history reveals origins of modern protein

synthesis. *PLoS ONE*, **7(3)**, e32776.
10. Caetano-Anollés, G. and Caetano-Anollés, D. (2015) Computing the origin and evolution of the ribosome from its structure–Uncovering processes of macromolecular accretion benefiting synthetic biology. *Comput. Struct. Biotech. J.*, **13**, 427-447.
11. Giraud, T., Pedersen, J.S. and Keller, L. (2003) Evolution of supercolonies: The Argentine ants of southern Europe. *Proc. Natl. Acad. Sci. USA*, **99**(9), 6075-6079.
12. Sunamura, E., Espadaler, X., Sakamoto, H., Suzuki, S., Terayama, M. and Tatsuki, S. (2009) Intercontinental union of Argentine ants: behavioral relationships among introduced populations in Europe, North America, and Asia. *Insectes Sociaux*, **56**(2), 143–147.
13. Bettencourt, L.M.A. (2013) The origin of scaling in cities. *Science* **340**, 1438-1441.
14. Schlomo, A. and Lamson-Hall, P. (2014) The rise and fall of Manhattan's densities, 1800-2010. Marron Institute of Urban Management, New York University. Working paper 14, pp. 1-48.
15. Hartwell, L.H., Hopfield, J.J., Leibler, S. and Murray, A.W. (1999) From molecular to modular cell biology. *Nature*, **401**, c47–c52.
16. Simon, H. (1962). The architecture of complexity. *Proc. Am. Phil. Soc.*, **106**, 467–482.
17. Caetano-Anollés, G. (2017) RubisCO and the search for biomolecular culprits of planetary change. *Bioessays* **39**(11), doi:10.1002/bies.201700174.
18. Sun, F.-J. and Caetano-Anollés, G. (2008) Transfer RNA and the origins of diversified life. *Sci. Prog.*, **91**(3), 265-284.
19. Mittenthal, J., Caetano-Anollés, D. and Caetano-Anollés, G. (2012) Biphasic patterns of diversification and the emergence of modules. *Front. Genet.* **3**, 147.
20. Bailor, M.H., Sun, X. and Al-Hashimi, H.M. (2010) Topology links RNA secondary structure with global conformation, dynamics, and adaptation. *Science*, **327**, 202–206.
21. Hyeon, C. and Thirumalai, D. (2012) Chain length determines the folding rates of RNA. *Biopys. J.*, **102**, L11–L13.
22. Fontana, W. (2002) Modeling 'evo-devo' with RNA. *Bioessays*, **24**, 1164–1177.
23. Sun, F.-J. and Caetano-Anollés, G. (2008) The origin and evolution of tRNA inferred from phylogenetic analysis of structure. *J. Mol. Evol.*, **66**, 21–35.
24. Caetano-Anollés, G. and Sun, F.-J. (2014) The natural history of transfer RNA and its interactions with the ribosome. *Front. Genet.*, **5**, 127.
25. Sun, F.-J. and Caetano-Anollés, G. (2010) The ancient history of the structure of ribonuclease P and the early origins of Archaea. *BMC Bioinformatics*, **11**, 153.
26. Sun, F.-J. and Caetano-Anollés, G. (2009) The evolutionary history of the structure of 5S ribosomal RNA. *J. Mol. Evol.*, **69**, 430–43.
27. Caetano-Anollés, G., Nasir, A., Zhou, K., Caetano-Anollés, D., Mittenthal, J.E., Sun, F.-J. and Kim, K.M. (2014) Archaea: The first domain of diversified life. *Archaea*, **2014**, 590214.
28. Tanaka, T. and Kikuchi, Y. (2001) Origin of the cloverleaf shape of transfer RNA—the double-hairpin model: implication for the role of tRNA intron and the long extra loop. *Viva Origino*, **29**, 134–142.
29. Wang, M., Jiang, Y.-Y., Kim, K.M., Qu, G., Ji, H.-F., Mittenthal, J.E., Zhang, H.Y. and Caetano-Anolles, G. (2011) A universal molecular clock of protein folds and its power in tracing the early history of aerobic metabolism and planet oxygenation. *Mol. Biol. Evol.*, **28**, 567–82.
30. Smirnov, A.V., Entelis, N.S., Krasheninnikov, I.A., Martin, R. and Tarassov, I.A. (2008) Specific features of 5S rRNA structure–Its interactions with macromolecules and possible functions. *Biochemistry*, **73**,1418–1437.
31. Ban, N., Nissen, P., Hansen, J., Moore, P.B. and Steitz, T.A. (2000) The complete atomic structure of the large ribosomal subunit at 2.4 Å resolution. *Science*, **289**, 905–920.
32. Caetano-Anollés, G., Wang, M. and Caetano-Anollés, D. (2013) Structural phylogenomics retrodicts the origin of the genetic code and uncovers the evolutionary impact of protein flexibility. *PLoS ONE*,

**8**, e72225.
33. Root-Bernstein, M., Root-Bernstein, R. (2015) The ribosome as a missing link in the evolution of life. *J. Theor. Biol.*, **367**, 130–158.
34. Farias, S.T., Rego, T.G. and José, M.V. (2014) Origin and evolution of the peptidyl transferase center from proto-tRNAs. *FEBS Open. Bio.*, **4**, 175-178.
35. Bloch, D., McArthur, B., Widdowson, R., Spector, D., Guimarães, R.C., Smith, J. (1984) tRNA-rRNA sequence homologies: A model for the origin of a common ancestral molecule, and prospects for its reconstruction. *Orig. Life.*, **14**, 571-578.
36. Caetano-Anollés, D. and Caetano-Anollés, G. (2016) Piecemeal buildup of the genetic code, ribosomes and genomes from primordial tRNA building blocks. *Life*, **6**, 43.
37. Caetano-Anollés, D. and Caetano-Anollés, G. (2017) Commentary: History of the ribosome and the origin of translation. *Front. Mol. Biosci.*, **3**, 87.
38. Eddy, S.R. (2011) Accelerated profile HMM searches. *PLoS Comput. Biol.*, **7**, e1002195.
39. Murzin, A., Brenner, S.E., Hubbard, T. and Clothia, C. (1995) SCOP: a structural classification of proteins for the investigation of sequences and structures. *J. Mol. Biol.*, **247**, 536–540.
40. Orengo, C.A., Michie, A., Jones, S., Jones, D.T., Swindells, M. and Thornton, J.M. (1997) CATH–a hierarchic classification of protein domain structures. *Structure*, **5**, 1093–1109.
41. Caetano-Anollés, G. and Caetano-Anollés, D. (2003) An evolutionary structured universe of protein architecture. *Genome Res.*, **13**, 1563-1571.
42. Nasir, A. and Caetano-Anollés, G. (2015) A phylogenomic data-driven exploration of viral origins and evolution. *Sci. Adv.*, **1**, e1500527.
43. Bukhari, S.A., and Caetano-Anollés, G. (2013) Origin and evolution of protein fold designs inferred from phylogenetic analysis of CATH domain structures in proteomes. *PLoS Comput. Biol.*, **9(3)**, e1003009.
44. Caetano-Anollés, G., Kim, H.-S., and Mittenthal, J.E. (2007) The origin of modern metabolic networks inferred from phylogenomic analysis of protein architecture. *Proc. Natl. Acad. Sci. USA*, **104**, 9358–9363.
45. Caetano-Anollés, K. and Caetano-Anollés, G. (2016) Structural phylogenomics reveals gradual evolutionary replacement of abiotic chemistries by protein enzymes in purine metabolism. *PLoS One*, **8(3)**, e59300.
46. Guimerà, R., and Amaral, L.A.N (2005) Functional cartography of complex metabolic networks. *Nature*, **433**, 895-900.
47. Barabási, A.-L. and Oltvai, Z.N. (2004) Network biology: understanding the cell's functional organization. *Nature Rev. Genet.,* **5**, 101–113.
48. Debès, C., Wang M., Caetano-Anollés, G. and Gräter, F. (2013) Evolutionary optimization of protein folding. *PLoS Comput. Biol.,* **9**, e1002861.
49. Goncearenco, A. and Berezovsky, I.N. (2015) Protein function from its emergence to diversity in contemporary proteins. *Phys. Biol.,* **12**, 045002.
50. Alva, V., Söding, J. and Lupas, A.N. (2015) A vocabulary of ancient peptides at the origin of folded proteins. *eLife*, 4, e09410.
51. Aziz, M.F., Caetano-Anollés, K. and Caetano-Anollés, G. (2016) The early history and emergence of molecular functions and modular scale-free network behavior. *Sci. Rep.,* **6**, 25058.
52. Wang, M. and Caetano-Anollés, G. (2009) The evolutionary mechanics of domain organization in proteomes and the rise of modularity in the protein world. *Structure,* **17**, 66–78.
53. Wang, M., Kurland, C.G. and Caetano-Anollés, G. (2011) Reductive evolution of proteomes and protein structures. *Proc. Natl. Acad. Sci. USA*, **108**(29), 11954-11958.
54. Nasir, A., Kim, K.M. and Caetano-Anollés, G. (2017) Phylogenetic tracings of proteome size support the gradual accretion of protein structural domains and the early origin of viruses from primordial cells. *Front. Microbiol.*, **8**, 1178.

55. Nasir, A., Kim, K.M. and Caetano-Anollés, G. (2014) Global patterns of protein domain gain and loss in superkingdoms. *PLoS Comput. Biol.*, **10**, e1003452.
56. Pietrzak, J. (2015) Shaping and structuring of high-rise office buildings in Europe. *Challenges of Modern Technology,* **6**(2), 48-56 31-38.
57. Janin, J. Bahadur, R.P. and Chakrabarti, P. (2008) Protein-protein interaction and quaternary structure. *Quaterly Rev. Biophys.*, **41**, 133-180.
57. Meyer, P.S., Yung, J.W. and Ausubel, J.H. (1999) A primer on logistic growth and substitution. The mathematics of the Loglet lab software. *Tech. Forecasting Social Change*, **61**, 247-271.
58. Grübler, A. (1970) The Rise and fall of infrastructures. Springer-Verlag, New York.
59. Rogers, E.M. (1962) Diffusion of innovations. Free Press, New York.
60. Bass, F. (1969) A new product growth for model consumer durables. *Management Sci.*, **15**(5), 215–227.
61. Caetano-Anollés, G., Yafremava, L. and Mittenthal, J.E. (2010) Modularity and dissipation in evolution of macromolecular structures, functions, and networks. In: Caetano-Anollés G (ed) Evolutionary bioinformatics and systems biology. Wiley-Blackwell, Hoboken, NJ, USA, pp. 443–449.

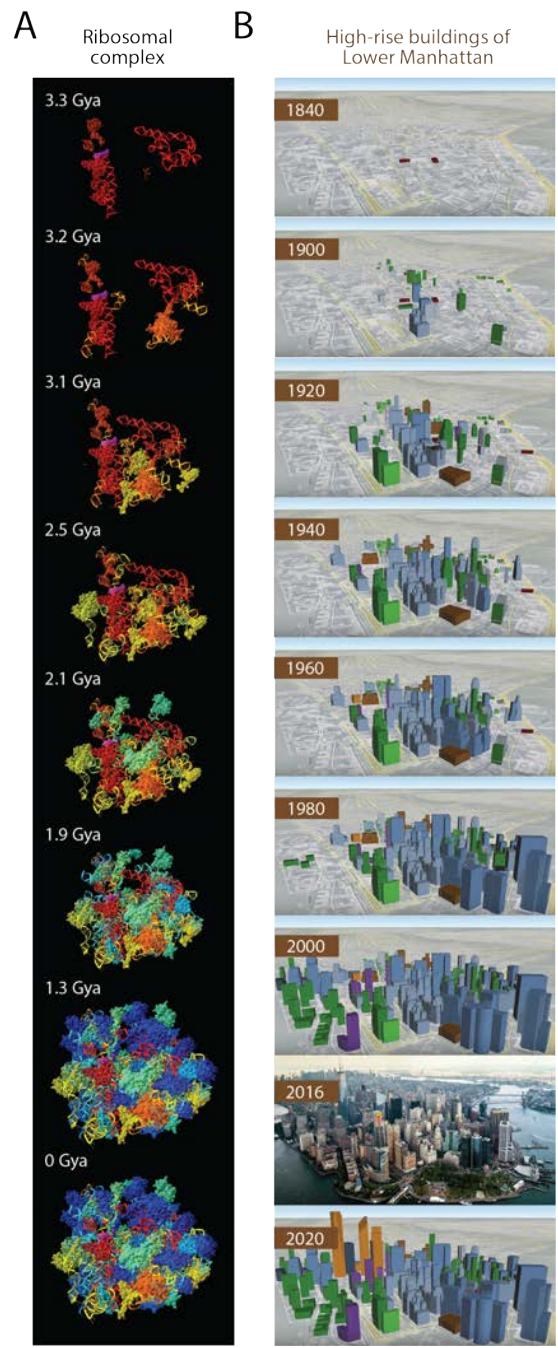

*Fig. 1. Accretion of RNA helices in the ribosomal complex and high-rise buildings in the financial district of the City of New York.* (**A**) *The natural history of the ribosome inferred from the phylogenomic analysis of protein structural domains and ribosomal RNA structures. RNA helical structures and structural domains of ribosomal proteins are colored according to their age in a scale from red (very ancient) to blue (very recent). Dates of the structural timeline are given in billions of years (Gy). Data from Harish and Caetano-Anollés[9].* (**B**) *Floor-level visualization of high-rise buildings in Lower Manhattan appearing since 1840 and predicted for 2020 (see animation from CubeLease, https://www.youtube.com/watch?v=Vc5r8osWauU).*

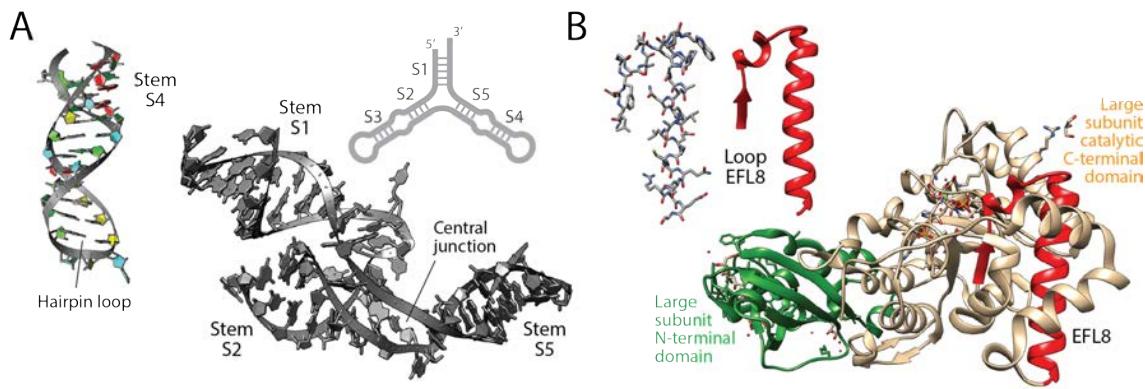

*Fig. 2.* The modular structure of functional nucleic acids and proteins. **A.** Stems are helical modules that make up RNA. The crystallographic model of stem S1 of the Escherichia coli 5S rRNA molecule (PDB entry 1c2x) on the left shows sugar/bases and backbone in fill/slab and ribbon representations, respectively, visualized using USFC Chimera. The central segment of the 5S rRNA molecule on the right shows a family C junction in which stem S1 bends towards stem S2 and stems S2 and S4 are stacked onto each other. The inset indexes the stem components in a secondary structure representation of the 5S rRNA molecule. **B.** A crystallographic model of the large subunit of the dimer making up the hexameric RubisCO enzyme from rice (PDB entry 1WDD) shows the typical 2-domain structure. The EFL8 loop of RubisCO (in red) is a Glycine-rich β-turn-α motif of the β/α-barrel fold that was recruited into the growing molecule ~3.5 billion years-ago (Gya) to provide redox functions[17]. The models of EFL8 on the left show details of the polypeptide chain and their associated secondary structures.

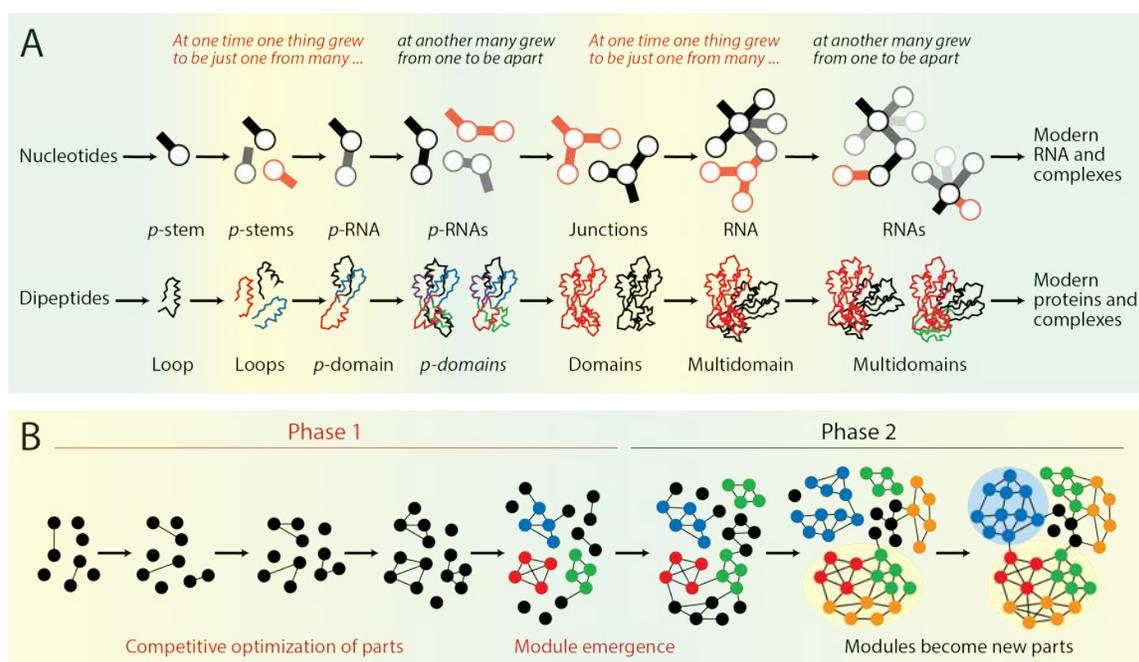

*Fig. 3.* The evolutionary growth of macromolecules by accretion of component parts leads to the modern world of proteins and functional RNA. **A.** The evolutionary growth of nucleic acids and proteins is explained by a 'double tale', one of accretion and the other of diversifying innovation. For nucleic acids, a first tale describes how nucleotides assembled into complex nucleic acids. First, they formed short polynucleotide strands that gained helical conformation through base pairings and stacking interactions. Condensations of these p-stems formed small hairpins (such as S4 of panel A), which acted as modules to form more complex 'junction' branching structures. RNA stems are illustrated with solid bars and loops with open circles. Similarly, a double tale describes the emergence of complex protein molecules. The first tale recounts how amino acids and dipeptides assembled into small peptide molecules through statistically biased condensations. Some of these peptides produced 'elementary' loops capable of interacting with ligands (such as EFL8 of panel B). Condensations of loops enhanced the accretion process by producing larger peptide structures with novel molecular functionalities. We name these structures proto-domains (p-domains). The second tale describes how p-domains diversified into a multitude of p-domains, which then diversified into domain families. The double tale is again revisited when it is observed how domains formed domain families, which combined with other domains to form multidomain proteins (such as the large subunit RubisCO dimer). These larger macromolecules diversified into modern proteins which then accreted to form the molecular complexes of the cell. Variant sequences and structures are illustrated with differently colored loop or domain backbones. **B.** A generic biphasic model of module creation illustrated with the emergence of network structure in evolution. Nodes of the network are parts of a growing system, and connections represent links of interaction. The larger number of links, the more cohesive and stable is the structure of a subnetwork. The rise of hierarchical modularity during phase 1 results in small highly connected subnetworks. These subnetworks become modules, which in phase 2 coalesce by combination into higher modules of network structure (highlighted with shades of yellow and blue). The model is inspired by the work of Mittenthal et al.[19].

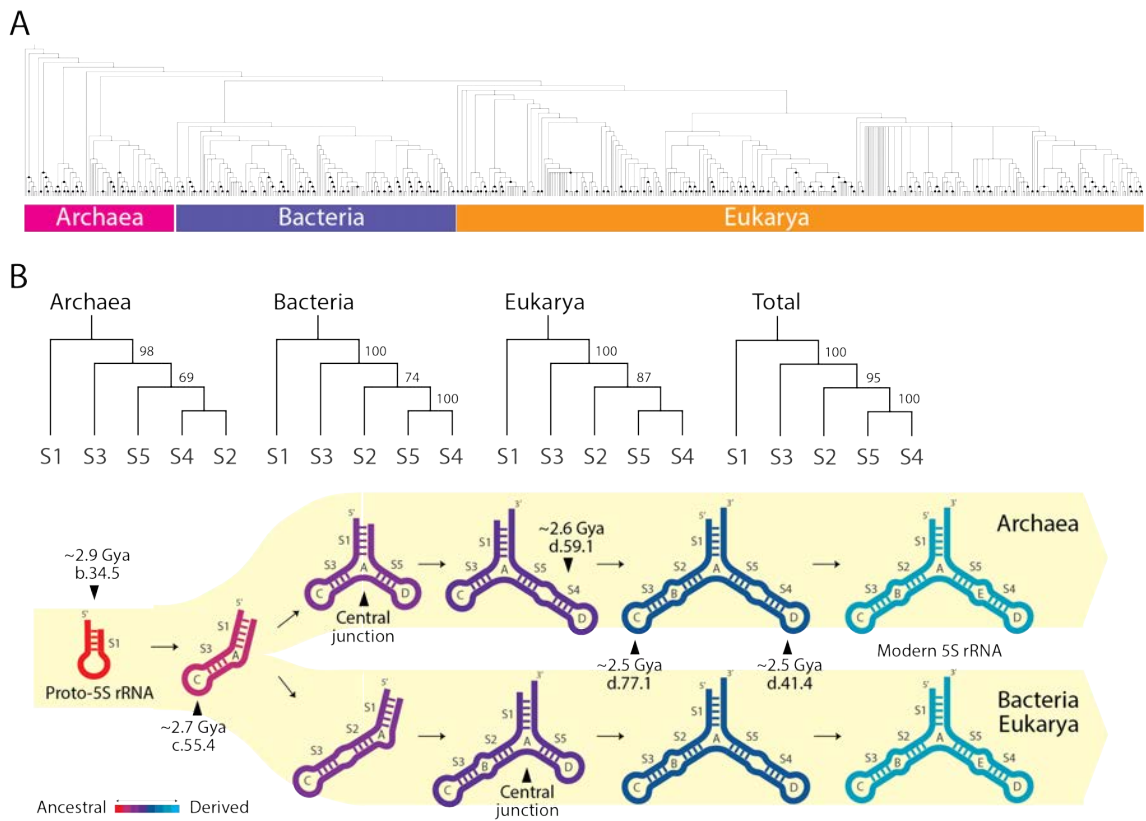

*Fig. 4. Evolution of 5S rRNA.* **A.** *A rooted tree of molecules (ToM) reconstructed from the sequence and structure of 665 5S rRNA molecules from all cellular superkingdoms of life shows the early rise of Archaea. Terminal leaves are not labeled since they would not be legible.* **B.** *A model of molecular evolution derived directly from rooted ToS reconstructions (top) that describes the accretion of helical stems in the evolution of the 5S rRNA molecules. Note the distinct evolutionary routes taken by ancestors of Archaea and ancestors of Bacteria and Eukarya, respectively, which match the overall topology of the ToM described in panel A. Also note the early formation of the central junction in the archaeal lineage. Arrowheads indicate interactions with r-proteins indexed with their SCOP name and evolutionary age. Modified from Sun and Caetano-Anollés[26].*

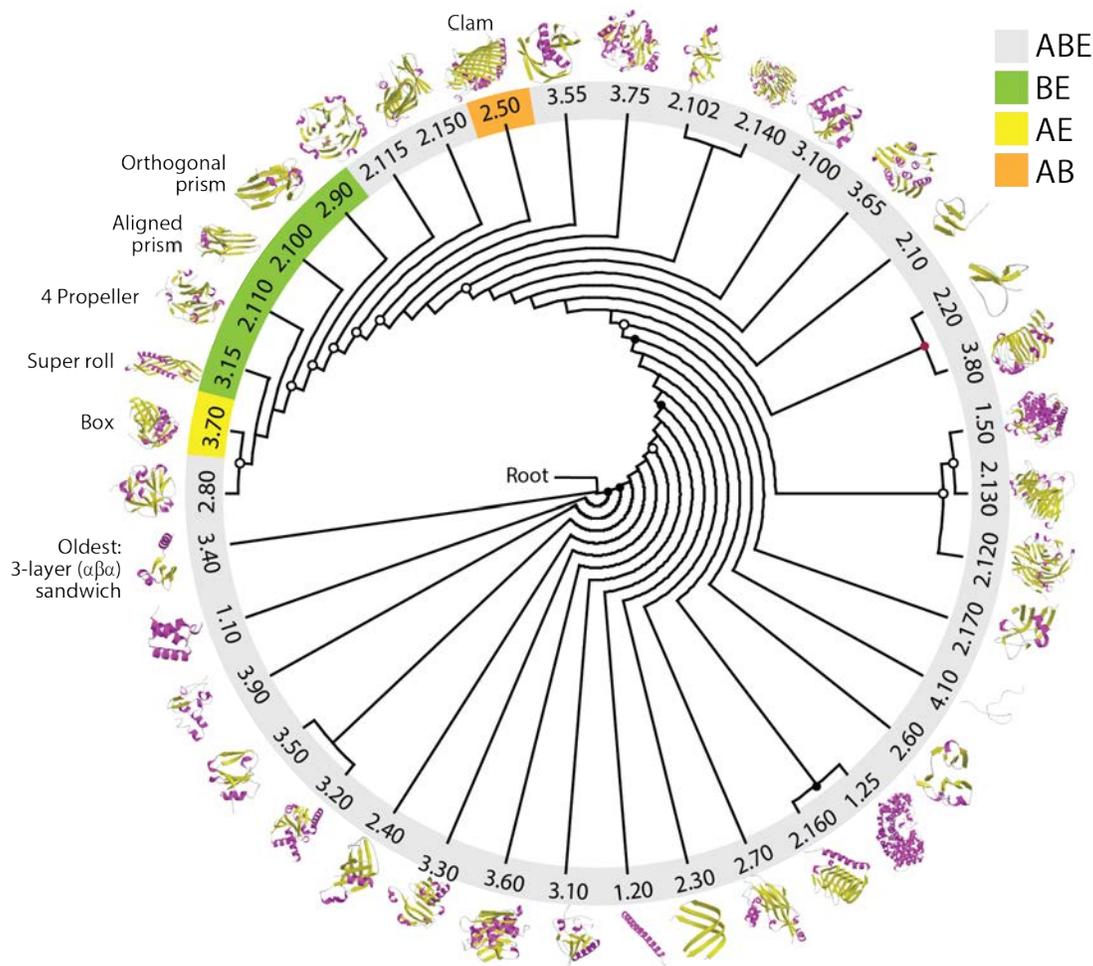

*Fig. 5. Phylogenomic tree of CATH 'architectures' describing the evolution of domain structures at a high level of structural abstraction. The phylogeny is a tree of domains (ToD) with leaves showing cartoon structural representations of CATH architectures with labels shaded according to their distribution in superkingdoms; ABE, shared by Archaea, Bacteria and Eukarya; BE, shared by Bacteria and Eukarya; AE, shared by Archaea and Eukarya; AB, shared by Archaea and Bacteria. Most architectures are universal and older than the rest. None are superkingdom-specific. Sandwiches and bundles are the oldest protein designs. Modified from Bukhari and Caetano-Anollés[43].*

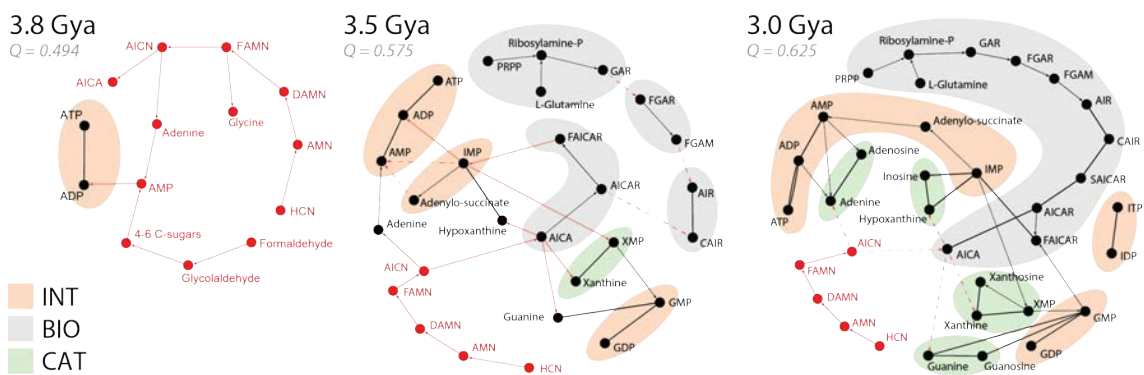

*Fig. 6.* *The emergence of the purine metabolic network. The reconstruction of metabolic subnetworks present 3.8, 3.5 and 3 billion years ago (Gya) reveal the piecemeal recruitment of functional modules for the nucleotide interconversion (INT), catabolism and salvage (CAT) and biosynthetic (BIO) pathways. Plausible metabolites and prebiotic chemistries supporting the emergent enzymatic reactions are depicted with red nodes and connections, respectively. Unknown reaction candidates or withering prebiotic pathways are indicated with dashed lines. Note how separate components unify into a cohesive network of INT, CAT and BIO modules. The network was rendered using the energy spring embedders and the Fruchterman-Reingold algorithm of Pajek. Full metabolite names can be found in Caetano-Anollés and Caetano-Anollés[45]. Modularity (Q) measures the density of connection in node communities. It increases in evolution.*

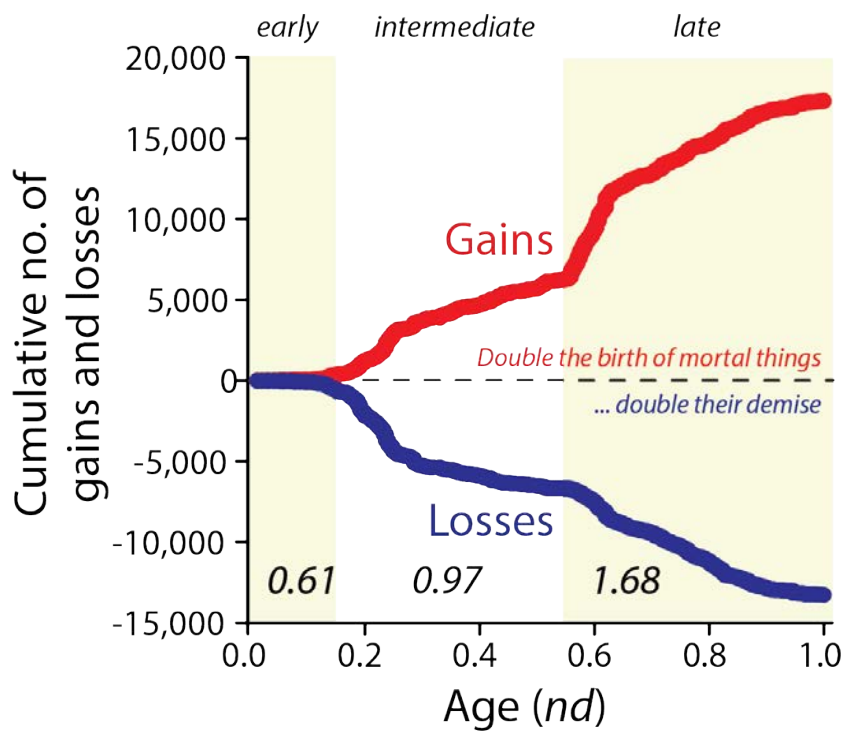

*Fig. 7.* Cumulative numbers of gains and losses in the occurrence of SCOP domain families along the branches of a tree of life. Scatter plots reveal growth trends in the accumulation of family gains and losses in the proteomes of 420 organisms belonging to Archaea (48), Bacteria (239) and Eukarya (133). Gains are identified in red while losses are identified in blue. The three evolutionary epochs of the protein world (early, intermediate, late) are marked with corresponding average gain-to-loss ratios in italics. A similar pattern is obtained when studying abundance of domain families. Data from Nasir et al.[55].

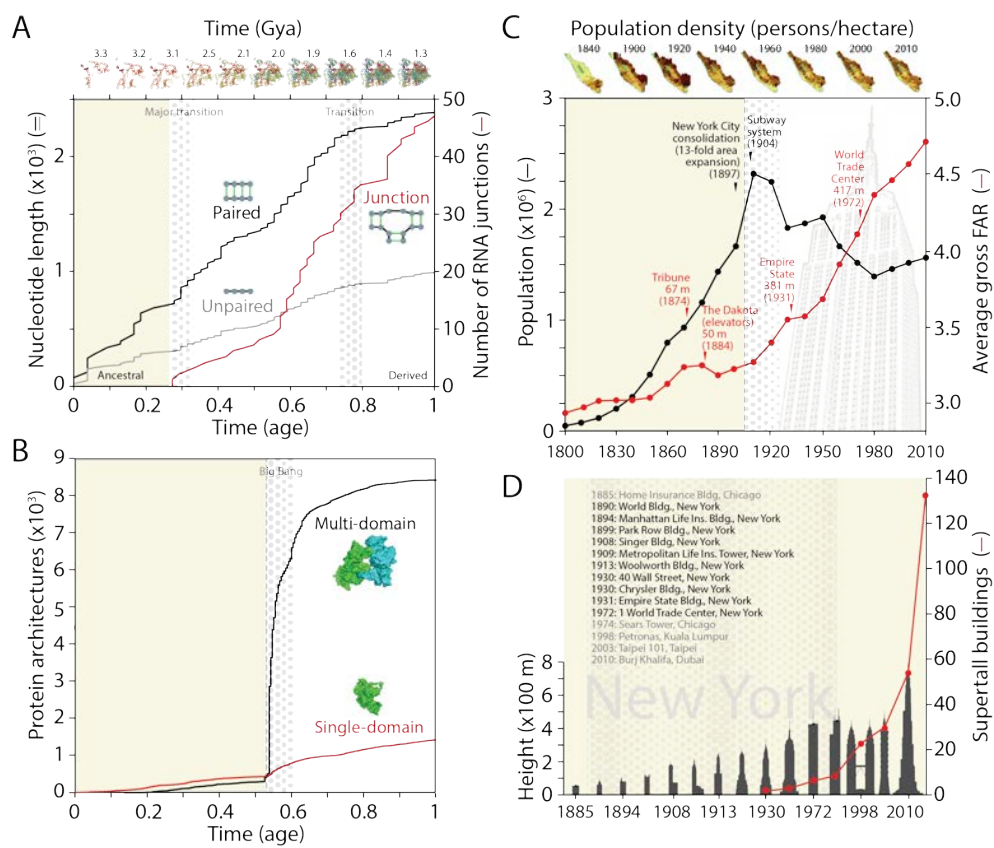

*Fig. 8. Emergence of modules in ribosomal RNA, proteins and the City of New York.* **A.** *The evolutionary accretion of nucleotides in ribosomal RNA in both paired (helical stems) and unpaired regions is described with cumulative plots of macromolecular growth. A plot of the first appearance of a 3-way RNA junction reveals that they emerge as a new kind of ribosomal module only after the 'major transition' in ribosomal evolution occurred. Note that junctions stabilize the complex ribosomal structure. The natural history of rRNA is shown on the top of the panel, with RNA structures colored according to their age in a scale from red (very ancient) to blue (very recent) and indexed with age (billions of years ago, Gya). Data from Harish and Caetano-Anollés[9].* **B.** *Cumulative plots describing the accumulation of single domain or multidomain proteins in evolution. Cumulative number is given as a function of time, with time described as an age of protein architecture in a relative 0-to-1 scale. The rise of multidomain proteins appears massive after a 'big bang' of domain combinations. Data from Wang and Caetano-Anollés[52].* **C.** *Growth of the human population of the Manhattan Island and the cumulative gross Floor Area Ratio (FAR) of its buildings. Corresponding bar plots of population density mapped onto a representation of the Manhattan island are provided on the top of the panel. Growth patterns show a biphasic behavior, with density peaking in 1910, then sharply declining for ~70 years, and finally slowly increasing until the present. The consolidation of the City of New York enables city expansion and a decrease of population density in the island. FAR is a common measure, commonly used when issuing building permits, that measures the ratio between the floor area of a building and the area of the plot on which it is built. Taller buildings have higher FAR. The graph shows that FAR levels increased with building growth, which correlates with high-rise building height. Data and visualizations were obtained from the NYU Stern Urbanization Project[14].* **C.** *History of the tallest buildings and the cumulative number of supertall buildings of the world (>300 m; March 2018). Data and building silhouettes from The Skyscraper Center (http://www.skyscrapercenter.com).*

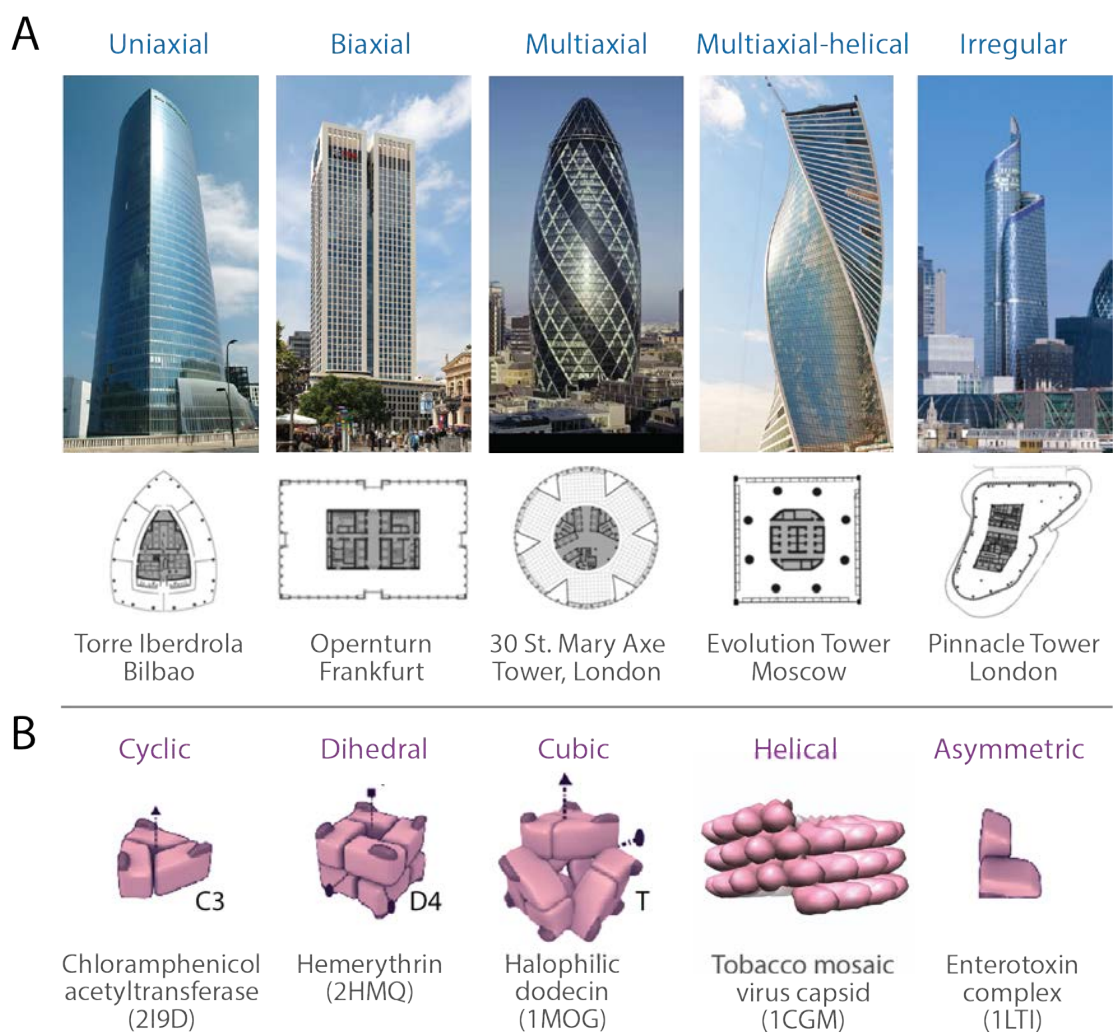

*Fig. 9.* *A systematics of symmetry describing spatial forms of skyscrapers and quaternary structure of protein complexes.* **A.** *Supertall buildings with a single inner core can be systematized into four groups according to the symmetry of their floor plan. Further examples of buildings and floor plans of this classification can be found in Pietrzak[56].* **B.** *Multimer symmetry of proteins is determined by imposing rotation, translation and other operations on the repeated structure. The Cyclic symmetry groups have one rotational axis of symmetry (C1, monomeric, C2, dimeric, etc.), which generally forms hollow tubes or directed shapes. The Dihedral symmetry groups contain one or more additional perpendicular axes of 2-fold symmetry (D1, D2, etc.) that expand the possible number of shapes in space. The Cubic symmetry groups (O, octahedral; T, tetrahedral, I, icosahedral) have 3-fold symmetry combined with non-perpendicular rotational axes. They are known to form protein cages. The Helical symmetry uses rotation and translation to create extended protein filaments. Asymmetric group arrangements result in irregular protein forms.*

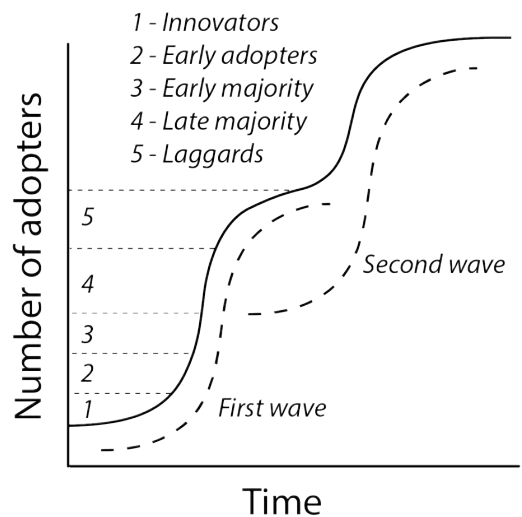

Eq. 1: $\frac{d\rho(t)}{dt} = (p + q\rho(t))(1 - \rho(t))$

Eq. 2: $d\rho(t) = p + (q - p)\rho(t) - q[\rho(t)]^2$

*Fig. 10.* *The discovery and growth of innovations described as a series of wavelets. A wavelet is a double S-curve that is typical of paths of high performance in diffusion of innovation models. These bi-logistic curves have loglets that often overlap in time with different magnitude and with different speeds generating sequential patterns of superposition, convergence and divergence[57]. The S-curve loglets can be modeled with Bass equations, which are listed below the plot.*